\documentclass[AMA,Times1COL]{WileyNJDv5} 

\usepackage[T1]{fontenc}
\usepackage[utf8]{inputenc}
\usepackage{fancyhdr}

\usepackage{times} 

\usepackage{colortbl}
\definecolor{HH}{RGB}{255, 0, 0}
\definecolor{HM}{RGB}{255, 85, 0}
\definecolor{HL}{RGB}{255, 170, 0}
\definecolor{MM}{RGB}{255, 255, 0}
\definecolor{ML}{RGB}{170, 255, 0}
\definecolor{LL}{RGB}{0, 255, 0}
\definecolor{HHP}{RGB}{0, 255, 0}
\definecolor{HMP}{RGB}{170, 255, 0}
\definecolor{MMP}{RGB}{255, 255, 0}
\definecolor{HLP}{RGB}{255, 170, 0}
\definecolor{MLP}{RGB}{255, 85, 0}
\definecolor{LLP}{RGB}{255, 0, 0}

\pdfoutput=1
\articletype{Article Type}%

\received{Date Month Year}
\revised{Date Month Year}
\accepted{Date Month Year}
\startpage{1}

\begin{document}

\title{Characterising Developer Sentiment in Software Components: An Exploratory Study of Gentoo}

\author[1]{Tien Rahayu Tulili}

\author[1]{Ayushi Rastogi}

\author[1]{Andrea Capiluppi}

\authormark{Tulili \textsc{et al.}}
\titlemark{Characterising Developer Sentiment in Software Components: An Exploratory Case Study of Gentoo}

\address[1]{\orgdiv{Bernoulli Institute for Mathematics, Computer Science and Artificial Intelligence}, \orgname{University of Groningen}, \orgaddress{\state{Groningen}, \country{The Netherlands}}}

\corres{Corresponding author Tien Rahayu Tulili, Andrea Capiluppi\email{t.r.tulili@rug.nl, a.capiluppi@rug.nl}}

\presentaddress{Nijenborgh 9, 9747AG, Groningen, Groningen, The Netherlands}


\abstract[Abstract]{Collaborative software development happens in teams, that cooperate on shared artefacts, and discuss development on online platforms. Due to the complexity of development and the variety of teams, software components often act as effective containers for parallel work and teams. 

Past research has shown how communication between team members, especially in an open-source environment, can become extremely toxic, and lead to members leaving the development team. This has a direct effect on the evolution and maintenance of the project in which the former members were active in. 

The purpose of our study is two-fold: first, we propose an approach to evaluate, at a finer granularity, the positive and negative emotions in the communication between developers; and second, we aim to characterise a project's development paths, or components, as more or less impacted by the emotions. 

Our analysis evaluates single sentences rather than whole messages as the finest granularity of communication. The previous study found that the high positivity or negativity at the sentence level may indirectly impact the writer him/herself, or the reader. In this way, we could highlight specific paths of Gentoo as the most affected by negative emotions, and show how negative emotions have evolved and changed along the same paths.

By joining the analysis of the mailing lists, from which we derive the sentiment of the developers, with the information derived from the development logs, we obtained a longitudinal picture of how development paths have been historically affected by positive or negative emotions. Our study shows that, in recent years, negative emotions have generally decreased in the communication between Gentoo developers. We also show how file paths, as collaborative software development artefacts, were more or less impacted by the emotions of the developers.
}


\jnlcitation{\cname{%
\author{T. R. Tulili},
\author{A. Rastogi}, and
\author{A. Capiluppi}}.
\ctitle{Characterising Developer Sentiment in Software Components: An Exploratory Case Study of Gentoo.} \cjournal{\it J of Software: Evolution and Process.} \cvol{xxxx;00(00):1--19}.}

\maketitle
\section{Introduction}\label{sec1:_Introduction}
Collaborative software development happens in teams that collaborate on shared artefacts and discuss development, most often on online platforms. Due to the complexity of development and the variety of teams, software components often act as effective containers for parallel work and teams~\cite{narduzzo2005role,langlois2008hackers}. These smaller and modular components may foster collaboration by means of specialisation in teams, reduced coordination overhead, simplified onboarding, clearer responsibilities allocation, efficient bug isolation, offering external contributions, and reuse of code~\cite{mockus2000case}. 

Besides working in parallel on the software components, the teams discuss issues, ideas, or other items related to the development. As dynamic communication occurs during the development cycle, it has been reported that developers often involve positive and negative emotions through their written communication~\cite{murgia2018exploratory,murgia2014developers}. Their expressions contain, at times, very strong negative expressions, and those may trigger toxic interactions among team members~\cite{Sharp2015ClosingADoor, Ranzhin2019Iruin}. 

From the literature on emotions within software development, there have been mixed results that at times support negative emotions as a catalyst of productivity, and other times show their downsides.
For example, Gachechiladze et al.'s~\cite {gachechiladze2017anger} study focused on detecting the emotions of developers working on OSS projects, arguing that anger can serve as a catalyst for tools to support collaborative software development. The work also highlighted the need for a finer-grained model to distinguish between anger directed towards self, others, and objects. On a different but complementary aspect, Huq et al.~\cite{Huq2020} explored the relationship between developer sentiment and software bugs by analysing GitHub commits: the study found that the commits introducing, preceding, or fixing bugs were significantly more negative and had a higher proportion of emotional messages.  

Other studies analysed the relationship between one or more of the software processes surrounding software development and the developer's sentiment: Robinson et al.~\cite{robinson2016developer} for instance, investigated how changes in the software development process impact developer sentiment and found that there is a correlation between a change in the software development routine and a change in developer sentiment. Additionally, Madampe et al.~\cite{madampe2020towards} focused on the Agile teams, and they found that the Agile teams exhibit their emotions during receiving, implementing, and delivering the requirement changes in software development. As an extreme case of negative communication between developers, team members have been reported leaving a development team~\cite{garcia2013role, Anonymous2014}, or developers who avoided joining one or several development communities~\cite{raman2020stress, StackOverflowBlog}.  

The case study described in the analysis conducted by Garcia et al.~\cite{garcia2013role} focused on the Gentoo Open Source Software (OSS) project\footnote{https://www.gentoo.org/}: it reported that one of the main contributors left the project suddenly, specifically due to personal contrasts and dissatisfaction with the social environment of the project as a whole. Moreover, following the departure of the key contributor, the Gentoo community never managed to attain similar levels of performance\cite{zanetti2013rise}. 

This study builds upon the work of Garcia et al.~\cite{garcia2013role} to explore how negative emotions affect the behaviour of software developers. Our paper aims to extend these findings by observing the evolution of the same Gentoo community over 23 years, providing a modular context to understand the role of emotions in software development. Furthermore, we believe that our study helps to understand developers' behaviour, particularly the emotional landscape within software development environments. This is crucial for ensuring the well-being and satisfaction of developers. In addition, this may help foster a more inclusive and supportive intervention.

Moreover, not only analysing the emotional traits of developers but also joining them to the software's architectural perspective, our intention is to observe sentiments, such as negative and positive emotions, and whether they might arise in specific parts of the code. We believe that investigating the granular level of software components may give different perspectives in identifying specific components within the project where emotions are prevalent.

The aim of our work is to assess, particularly at a finer granularity,  the positive and negative emotions in the communication between developers; furthermore, we characterise a project's development paths or components as more or less impacted by the emotions. We define our research questions as follows:

\textbf{RQ1:}~\textbf{What is the extent of negative and positive emotions in the Gentoo developers' written communication?}
\\\textit{Rationale.} This question aims to trace the level of positive and negative communication used by Gentoo developers. This is done to complement and update the earlier work on Gentoo, performed 11 years ago in~\cite{garcia2013role}. Furthermore, it will investigate the emotional dynamics of the open source community, particularly in the Gentoo community, for over 23 years.

\textbf{RQ2:}~\textbf{How pervasive are the negative and positive emotions in Gentoo's components, and how did they evolve?}
\\\textit{Rationale.} This question aims to characterise the Gentoo components by quantifying the relative positive and negative communication in each. By acknowledging the characteristics of (less or more) both positive and negative components, we also observed how the levels of sentiment evolved in time in the same component. Identifying the changes may highlight important points about which components may be affected by more negative emotions. 

\textbf{RQ3:~How do the emotions affect Gentoo's developers' performance?}
\\\textit{Rationale.} This question aims to analyse further the possible effect of negative/positive paths on the developers' activity, in particular investigating whether the negativity or positivity may lead to the decrease of the performance done by developers in the components or otherwise.

This paper is articulated as follows: Section~\ref{sec:_relworks} surveys the literature and related works; Section~\ref{sec:_method} illustrates the methodology we used to conduct this case study, and Section~\ref{sec:_results} summarises the results. Section~\ref{sec:_discussion_and_implications} proposes some points for discussion and implications. Section \ref{sec:_threats to validity} illustrates the threats to validity, while Section~\ref{sec:_conclusion} concludes.

\section{Related Works}\label{sec:_relworks}
Our work is multidisciplinary in nature; therefore, we explored various avenues to cover the background work needed to underpin our study. We group the related works into four large topics below.

\textbf{Software components as collaboration and coordination blocks} --Wu et al.~\cite{wu2023social} investigated the role of social and technical dependency networks in OSS communities. They considered network metrics (e.g. degree centrality, betweenness centrality, and closeness centrality) in both module networks and developer networks to understand their influence on project success. Their findings were that nonlinear relationships exist between degree centrality and the number of connections in both social and technical networks and OSS success. In other words, changes in the number of connections an individual or a component has with others do not have a consistent effect on the success of OSS projects. Furthermore, Tee et al.~\cite{tee2019modular} delved into the challenges of collaboration in large-scale projects and examined the impact of modular designs on enhancing coordination. The authors explored how modularity can overcome coordination obstacles and hinder collaboration by emphasising specialisation within modules. They suggested that in the context of coordination and cooperation, there is a mutual interaction between modular designs and the adoption of integrating practices.

\textbf{Emotion states and developers' activity} --Several studies investigated how the developers' emotions affect their contribution to a project. Unhappiness may (e.g., emotions and mood) cause low productivity and work withdrawal~\cite{graziotin2018happens} or may lead the developers to leave the project~\cite{garcia2013role}. In addition, a study done by Huq et al.~\cite{Huq2020} observed the relationship between developer's emotions and software collaborative artefacts (e.g. commits) retrieved from GitHub repositories. The authors classified the commit's messages with the Senti4SD tool and found that bug-related commits contained more negative expressions than regular commits. Moreover, a study conducted by Madampe et al.~\cite{madampe2020towards} examined the emotional responses to requirement changes by agile teams. They found that agile teams show their emotions during software development and emphasised the importance of considering the emotional responses of agile teams when applying agile principles and practices to deal with changes.

When researchers attempted to detect the developers' negative emotions, a number of studies proposing emotion detection classifiers existed. Building a negative-emotion classifier in the developer's communication was done by Gachechiladze et al.~\cite{gachechiladze2017anger}. The authors focused on detecting the developer's angry emotion in three levels direction: own-self, others, and objects. Similar studies applying state-of-the-art sentiment tools or machine learning algorithms also was done in detecting  anger~\cite{ortu2016arsonists,murgia2018exploratory}. Furthermore, several studies observed other emotional responses, including sadness~\cite{wang2019emotions}, fear~\cite{murgia2018exploratory,wang2019emotions}, and stress~\cite{kim2018human}. They developed models that can detect those emotions or categorise the content of comments into certain emotions. Additionally, several studies focused on detecting arousal and valence~\cite{islam2018deva, islam2019marvalous} (emotional dimensions). They built machine-learning-based tools that can detect individual emotional states through his/her textual communication.

\textbf{Impacts of emotions on collaboration and team dynamics.} --Few studies investigated the influence of emotions on collaboration and team dynamics in general. For example, Das~\cite{das2023aligning} explored the alignment of emotions, thoughts, and feelings and their impact on team collaboration. In their quantitative study, they found that there exists a strong positive correlation between psychological safety and team effectiveness. They suggested that by developing emotional intelligence skills and utilising neuro-linguistic programming techniques, teams can enhance communication, collaboration, and problem-solving. Furthermore, more positive emotions have the potential to inspire team members to enhance their coordination, leading to improved team performance; conversely, more negative emotions can discourage members and downgrade them towards lower standards and lessen team performance~\cite{rafaeli2009sensemaking}. In addition, Kazemitabar~\cite{kazemitabar2024examining} examined the correlation between socially-shared emotion regulation (SSER) and team dynamics and interaction. The significant association between SSER and team dynamics and interactions highlighted the power of emotions in building key team coordination mechanisms and strengthening team performance. 

\textbf{Evolution of sentiment in software development} --Werder\cite{werder2018evolution} focused their longitudinal study on emotions in the open-source development teams of GitHub projects. They utilised the Syuzhet R package to extract positive and negative emotional displays; they analysed the developers' positive emotions during the development phase of one to one hundred and twenty months with Growth curve analysis. They found a decreasing linear pattern, suggesting that instances of positive emotional expressions decrease as time progresses. Furthermore, a study conducted by Robinson et al.,~\cite{robinson2016developer} investigated the correlation between the developer's behaviour and the sentiment in open-source projects from Github during the development over 66 weeks. In their variance model, they found a relationship between changes in behaviour, in particular in routine changes and changes in sentiment.

\vspace{0.2cm}
Despite all the literature mentioned, the investigation focusing on shared software artefacts with the perspective of sentiment analysis has so far attracted very little attention. As far as we have gathered in the analysis of related work, ours is the first study aimed at characterising the developers' sentiment in the software components as units of development. These units are also part of the projects or locations where OSS developers collaborated. In this work, we consider components as the artefacts, as these are the shared locations where all the projects' packages are built. We further analysed the discussion made during the development of the packages, in particular, looking at the negative emotions conveyed in the written communication.

\section{Methodology}\label{sec:_method}
In this section, we describe the methodology of our study and its steps, together with the definitions that were used. In our study, we explicitly selected a large open-source ecosystem, the Gentoo community, as the project to be analysed, since some interesting aspects of negative communication influencing developers' behaviour emerged in the past~\cite{garcia2013role}. The Gentoo Linux project was established in 1999, and it still exists. The project mainly develops a free operating system based on Linux\footnote{https://www.gentoo.org/get-started/about/}. This huge project stays dedicated to its fundamental values of customisation, optimisation, and fostering community involvement in development.

From its inception, Gentoo has used its own mailing lists as one of the means of communication during software development. We retrieved the mailing lists from the online repository\footnote{https://archives.gentoo.org/gentoo-dev/} dating back to 2001, and the analysis has been performed for the years between 2001 and 2023. In the same time span, we considered the developers' commits and retrieved the data from the commits repository\footnote{https://gitweb.gentoo.org/repo/gentoo/historical.git/, https://github.com/gentoo/gentoo.git}. 

\subsection{Definitions}\label{sec:_methods_definitions} 
In our study, we make use of the following definitions: 
\begin{itemize}
\item a \textit{\textbf{commit}} is a revision of one or more files made by one or more developers; 
\item a \textit{\textbf{head path}} (or simply `path') is the high-level location of a file within the code base: for example, the ‘\textit{media-libs/id3lib/files/digest-id3lib-3.8.0\textunderscore pre2}’ file belongs to the \textit{`media-libs'} head path;

\item a \textit{\textbf{path grain}} is the high-level name of each path, connected to a larger component: as an example, in Gentoo there are several paths such as  `\textit{media-libs}', `\textit{media-gfx}', `\textit{media-sound}', `\textit{media-video}', `\textit{media-plugins}', `\textit{media-fonts}', `\textit{media-tv}', `\textit{media-radio}' paths points to the \textit{`media'} grain. This is a necessary adjustment for the Gentoo code base specifically, as we noticed that several paths refer to the same larger component; 

\item an \textit{\textbf{‘active developer’}} refers to the developers pushing commits into the Gentoo repository. 

\item a \textit{\textbf{negative message}} refers to a message containing negative sentences with negative scores ranging between -3 and -5 and positive scores ranging between +1 and +2. The range $[-3; -5]$ contained more strong negative words/phrases such as `really hate', `very afraid', `awful', `sad', `suffer', `dislike', and so on. 

\item a \textit{\textbf{positive message}} refers to a message containing positive sentences with positive scores ranging between +3 and +5 and negative scores ranging between -1 and -2. The range $[+3; +5]$ contained more strong positive words/phrases such as `extremely funny', `very cool', `excited, `excellent', `thanks', and so on. 

\item a \textit{\textbf{developer writing negative sentences (DWNs)}} refers to developers writing negative messages with certain conditions. To evaluate the threshold for which a developer should be classified as DWN, we ran an analysis of all Gentoo developers (active or retired) and evaluated the number of negative messages for each. We considered as DWNs those developers belonging to the first quartile, i.e., the top 5\% of the distribution in the number of negative messages.

\item a \textit{\textbf{developer writing positive sentences (DWPs)}} refers to developers writing positive messages with certain conditions. The conditions to consider a developer a DWP are similar to the DWN above, but considering positive messages.
\end{itemize}

\subsection{Extracting mailing list and commits data}
The two datasets, mailing lists and commits, were collected separately. On the one hand, we scraped the openly available mailing list archive dating between January 2001 and March 2023, storing all the data in files. The \textit{metadata} of the mailing lists used in the study consists of \textit{‘From (Name)’,} \textit{‘From (Email)’}, \textit{‘To’}, \textit{‘CC’}, \textit{‘Subject’}, \textit{‘Date’}, \textit{‘Message-Id}’, \textit{‘Reply to’}, \textit{‘Message-Body’}. We collected all the messages, resulting in 127,584 emails. We stored the processed emails in the database. 

On the other hand, and in order to evaluate the activity of developers in Gentoo's paths, we used the command ‘\textit{git clone}’ to retrieve all the commits dating from July 2000 and March 2023 from the historical and GitHub repository, drilled the data with PyDrill, and stored it in the formatted files. We stored the files' metadata in columns of a database for our analysis. The metadata of commits data consists of \textit{‘hash’}, \textit{‘author\textunderscore email’}, \textit{‘author\textunderscore name’}, \textit{‘author\textunderscore date’}, \textit{‘committer’}, \textit{‘committer\textunderscore date’}, \textit{‘filename’}, \textit{‘old \textunderscore path’}, and \textit{‘new\textunderscore path’}.

In total, we collected 5,500,779 commits for our analysis.

\subsection{Pre-processing the data}
 In this phase, we cleaned up the content of the message body. We removed lines prefixed by the character ‘\textgreater’, URLs, name(s) or signatures, and greetings (i.e. “Kind Regards”, and “Best Regards”). In addition, we removed lines containing code syntax and/or HTML/XML tags. 
 
As in the mailing list, some of the developers have more than one email address; we normalised those emails by choosing only one email address and making them uniform to the same developer. This follows the same approach as shown in~\cite{von2003community}. Furthermore, if the email does not have a name and only has an email address, we normalised the name by extracting the string derived from the account name of the email address. 
 
We pre-processed the name of the component file path for each commit to obtain both the \textit{‘head path’} and the \textit{‘path grain’} as defined above. These granular path names helped us identify which developers collaborated on what paths or path grains. 

\subsection{Identifying sentiment in the mailing list}
In order to classify the mailing list messages, we employed the Sentistrength-SE tool~\cite{islam2018sentistrength}. Sentistrength-SE is an off-the-shelf sentiment analysis tool specifically tailored for Software Engineering and used several times by other researchers~\cite{calefato2018sentiment, Pletea2014, chen2021emoji, el2019empirical, girardi2021emotions}. The tool gives two values ( positive and negative values) on the text (a sentence) provided as input. The positive values fallin the $[+1;-+]$ range , and the negative values fall in between $[-1;+-]$.

When one considers email messages, those are typically made of many sentences, each holding a different sentiment score. We believe that evaluating a whole message as a unit of sentiment would risk watering down highly positive or highly negative sentences into one final (email-wide) score. As pointed out in~\cite{richter2010words}, high positivity or negativity at the sentence level may indirectly impact the writer him/herself, or the reader. 


As an example of our reasoning, the following snippet of a message\footnote{The message does not contain personal data, so it does not infringe the GDPR. In addition, the mailing list messages are publicly accessible and have been made public by the Gentoo community.}: `\textit{Wow. Your second stab at my team in 3 days, without me even responding. I should probably be blushing if it weren't for the fact that I really don't give a damn about you or anything that you say. Quite honestly, the same goes for pretty much anybody who works with you. You are a poisonous person to Gentoo and I sincerely wish that some day people around here will grow a pair and realise that your incessant self-absorbed bullshit simply isn't something we really want around here. I mean, we've already thrown out you and three of your cronies because your attitude sucks and you're all a pain in the ass to work with. What exactly do we need to do here? Ban you all? I find it massively amusing that most of the traffic on this list over the past 3 days has come from people that have been *FORCIBLY* remove from the Gentoo project. Oh yeah, don't bother responding to me. I've decided to put you and all of your little cohorts into my killfile so I no longer have to read your constant barrage of b***s**t. Seriously, you're a complete f***ing waste.}' 

The message as a whole contains 12 sentences: their negative scores are (-1, -2, -1, -1, -4, -4, -1, -1, -1, -1, -5, -5), while their positive scores are (4, 
 1, 1, 1, 1, 1, 1, 1, 1, 1, 1, 1). If that message was evaluated as a whole, its overall \textbf{average} negative score would be 2.25, and its overall \textbf{average} positive score would be 1.25 instead, signalling only mild negativity and low positivity and missing the presence of high negative sentences and highly positive sentences. 

In our approach, we decided to split all the email messages into \textit{sentences} and analyzed the sentiment at that level, focusing on both the positive and negative sentences. Our proxy was that each sentence should be terminated with a full stop (`.') or another character that indicates the end of the sentence (e.g., `?', `! ').

Additionally, regarding the annotations of negative, neutral, and positive labels on the messages on the sentence level done by two annotators, we conducted the annotating separately on 100 samples retrieved randomly from 100 negative sentences and 100 positive sentences. Each annotator gave positive (with a range between +1 and +5) and negative (with a range between -1 and -5) scores on each sentence.
As each sentence has two scores, after discussion, the annotators agreed to determine a formula to set one annotation, either positive, negative, or neutral, on each sentence. Subsequently, the Cohen Kappa was evaluated. We evaluated our results' annotation with the Cohen Kappa and have a moderate value \textit{0.507}.

\subsection{Identifying DWNs and DWPs}
Considering the pool of Gentoo developers, and in order to classify one as DWN, we obtained the first quartile (Q1) of the distribution. We found a value of 17, the minimum number of negative sentences in one year. This value represents the top 5\% of all senders writing negative sentences in the mailing list, which accounts for 164 developers.

From the pool of all official Gentoo developers (either active or retired), we also evaluated the DWPs based on the distribution of positive messages. The Q1 value of the distribution considers a developer to be a DWP if they had at least written 11 positive sentences in one year. As mentioned above, this number was chosen as the top 5\% of developers writing positive sentences on the mailing list. This threshold helped us isolate a pool of 174 DWPs.

\subsection{Aggregating, linking and visualising data for analysis}
We conducted different aggregations to scrutinise our datasets (e.g. mailing list and commits), particularly investigating negative emotions in the Gentoo mailing list. For the mailing list, we aggregated the number of positive and negative scores labelled by the Sentistrength-SE tool to obtain the overall picture of the sentiment trend throughout the period (between 2001 and 2023). We further observed both the positive and negative sentences, particularly at the sentence level, by considering scores $[+3; +5]$ and $[-3; -5]$, respectively. The reason we chose the ranges was explained in the Methodology section under the subsection `Definitions' and `Identifying sentiment in the mailing list'. We aggregated those positive and negative sentences separately by categories (e.g. developers and time - year, month, day). Similarly, we aggregated the negative and positive messages separately. 

For the component analysis, we considered that each commit has its own file path name. From that, we obtained the (head) \textit{path} and the \textit{grain}. Subsequently, we explored the commit data on the paths and grains, producing aggregations to investigate the commits done by each developer in each path and grain. 

Finally, we combined the results derived from the two datasets and used these results to produce visualisations. We linked the aggregation sets by date (year, month, and day). We selected variable date as the joined variable as we believe that the date the developers built the components may be similar to the date they discussed the components they had been developing.

Lastly, we visualised the results into the necessary graphs to address our research questions. During the visualisation process, we conducted normalisation when necessary. For example, we have performed the normalisation on the number of differences between positive and negative messages to acknowledge any patterns across all the components over the period. We applied the z-score standard normalisation provided by R. 

\section{Results}\label{sec:_results}
\subsection{\textbf{RQ1 -- Extent of negative and positive emotions in the written communication in Gentoo}}

\begin{figure}
    \centering
    \includegraphics[width=.95\linewidth]{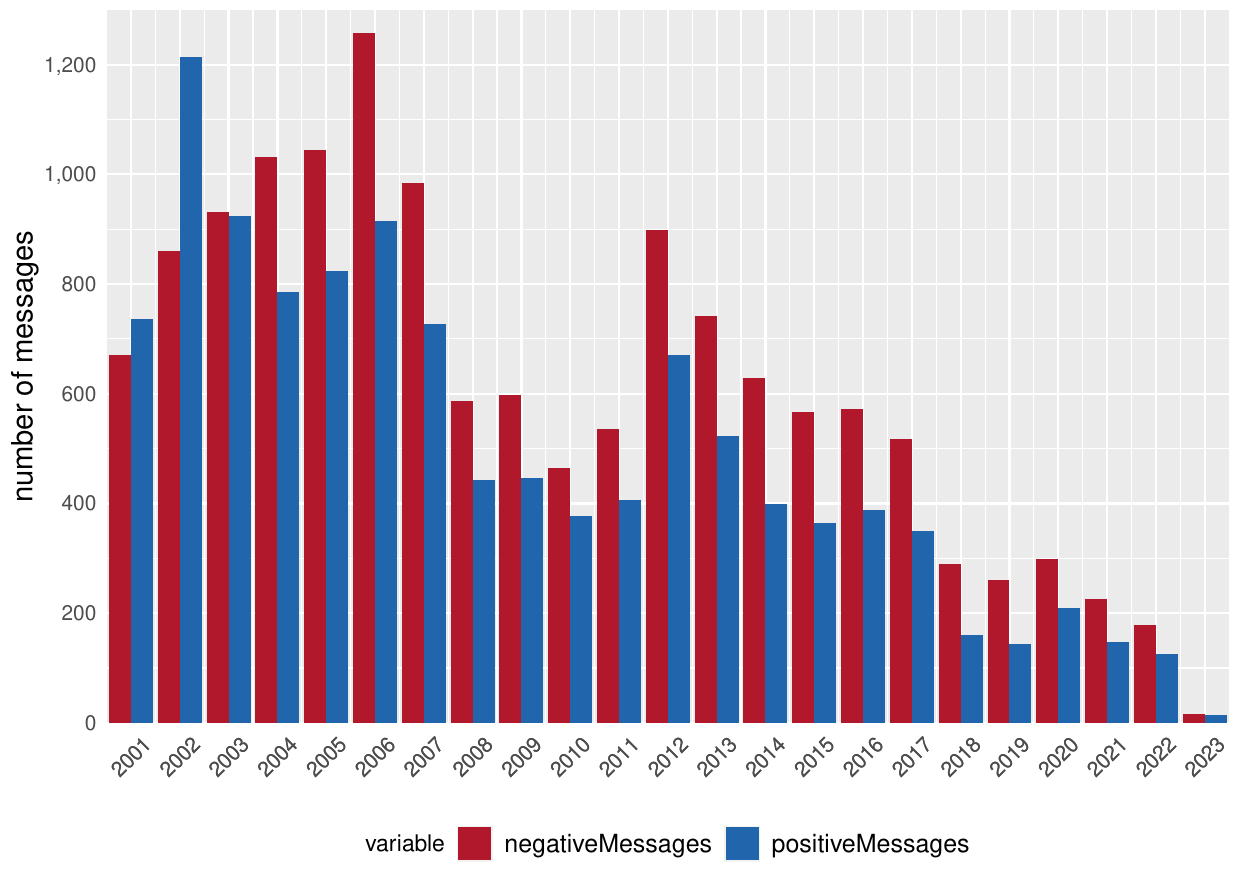}
    \caption{Total number of messages containing positive sentences (blue), and messages containing negative sentences (red).}
    \label{fig_RQ1:_histogram}
\end{figure}

Figure \ref{fig_RQ1:_histogram} shows the number of positive and negative messages per year (from 2001 to 2023). As per the definition above, the ‘\textit{negative messages}’ are those messages that contain at least one sentence with a negative score between -3 and -5, and the ‘\textit{positive messages}’ are those containing at least one sentence with a positive score between +3 and +5. 

In general, both positive and negative sentiments in messages can be detected during the Gentoo software development cycle of over two decades. We noticed that the emotional sentiments reached their three peaks in the absolute number of messages, which were in 2002, 2006, and 2012. It is furthermore evident that the trend of both sentiments has evolved and has experienced a decrease in intensity in the last few years. In further detail, the number of messages reached under 500 messages between 2018 and 2023. 

Although the total number of both messages decreased in recent years, the proportion of negative messages is greater than that of positive messages. At a glance, the bars of all the negative messages clearly showed a larger proportion than the positive ones. In detail, as of 2004, this proportion of negative messages was getting wider from the positive ones around 25\% to around 30\% in recent years.

\begin{figure}
    \centering
    {\includegraphics[width=.45\linewidth]{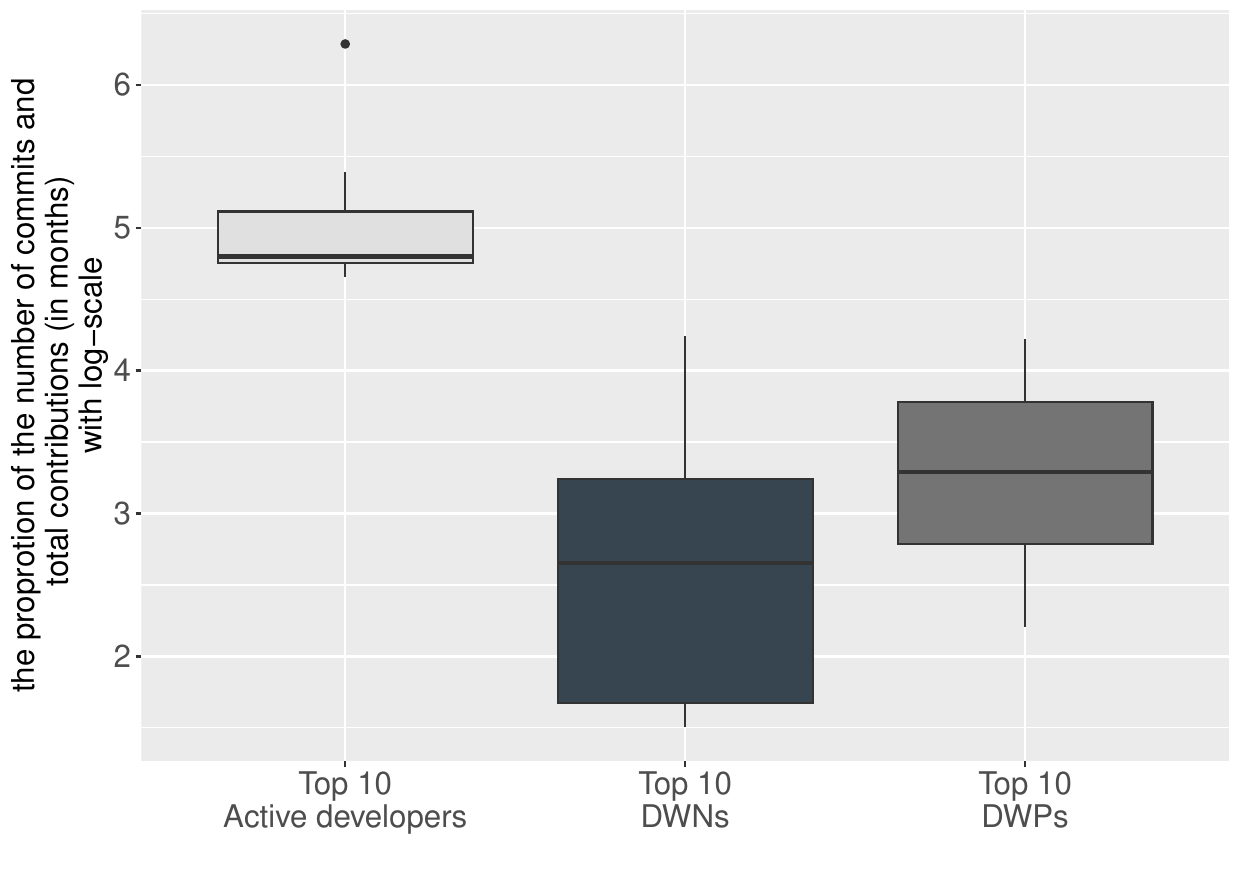}}
    \qquad
    {\includegraphics[width=.45\linewidth]{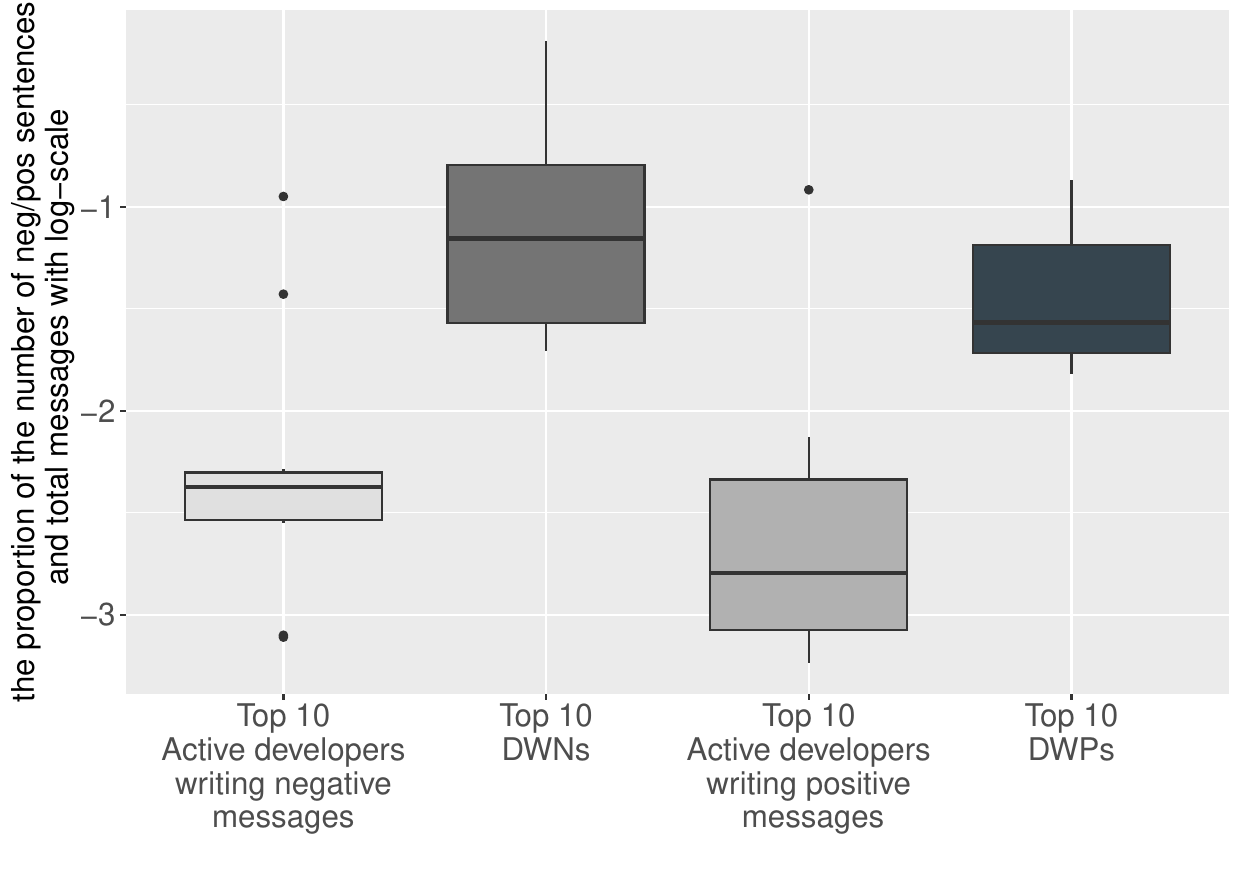}}
    \caption{Boxplots of the top 10 of active developers, DWNs, DWPs regarding a relative number of commits done (a) and of positive and negative sentences written (b). } 
    \label{fig_RQ1:_boxplots comparison} 
\end{figure}

On the one hand, having observed the yearly trends of negative and positive messages, we subsequently focused on the developers who reportedly sent most of those kinds of messages. Out of the overall number of Gentoo developers, we observed 164 developers who had ever sent at least 17 strong negative sentences in one year. (i.e., DWNs) during our observation period. Additionaly, we further observed 174 developers who had ever sent at least 11 strong positive sentences in one year (i.e., DWPs). 

On the other hand, we investigated the active developers, taking into account their monthly contributions. Subsequently, we compared this group with the other two groups (DWNs and DWPs) that were making the commits.

The most prolific DWNs and DWPs are not also the most active: comparing the top 10 DWNs and DWPs with the top 10 most active developers (i.e. those responsible for the most commits) shows significant differences. Figure~\ref{fig_RQ1:_boxplots comparison} shows the boxplots of the three groups in terms of the number of commits/working duration in months (a) and negative and positive messages/total messages written (b). The very-dark-grey-boxplot represents the top ten DWNs group, the dark-grey-boxplot represents the top ten 10 DWPs, and the light-grey-boxplot represents the top ten active developers.

From those two figures, we notice that the most active developers (Top 10 Active developers writing negative messages)  tend to use fewer negative sentences when communicating with other developers during software development. We statistically calculated the Wilcoxon of those two groups and ascertained the $p-value = 0.002807$. Hence, we reject the null hypotheses (“the most active developers and the DWNs contributed an equal amount of the proportion between the negative messages and total messages”). Furthermore, considering the DWNs, almost all of them committed much less than the active group but contributed more negative sentences. 

Moreover, we subsequently noticed that the most active developers (Top 10 Active developers writing positive messages) tend to use fewer positive messages than those of DWPs. We also detected statistically significant differences between those two groups: utilizing the Wilcoxon test, with a $p-value = 0.001932$, we rejected the null hypothesis: "the most active developers and the DWPs contributed an equal amount of the proportion between positive messages and total messages". 

\subsection{\textbf{Results: RQ2 -- Extent of negative and positive communication in Gentoo components}\label{sec:_results_RQ2a_extent_emotions_on_components}}
\begin{figure*}[ht]
    \centering
    \includegraphics[width=.7\linewidth]{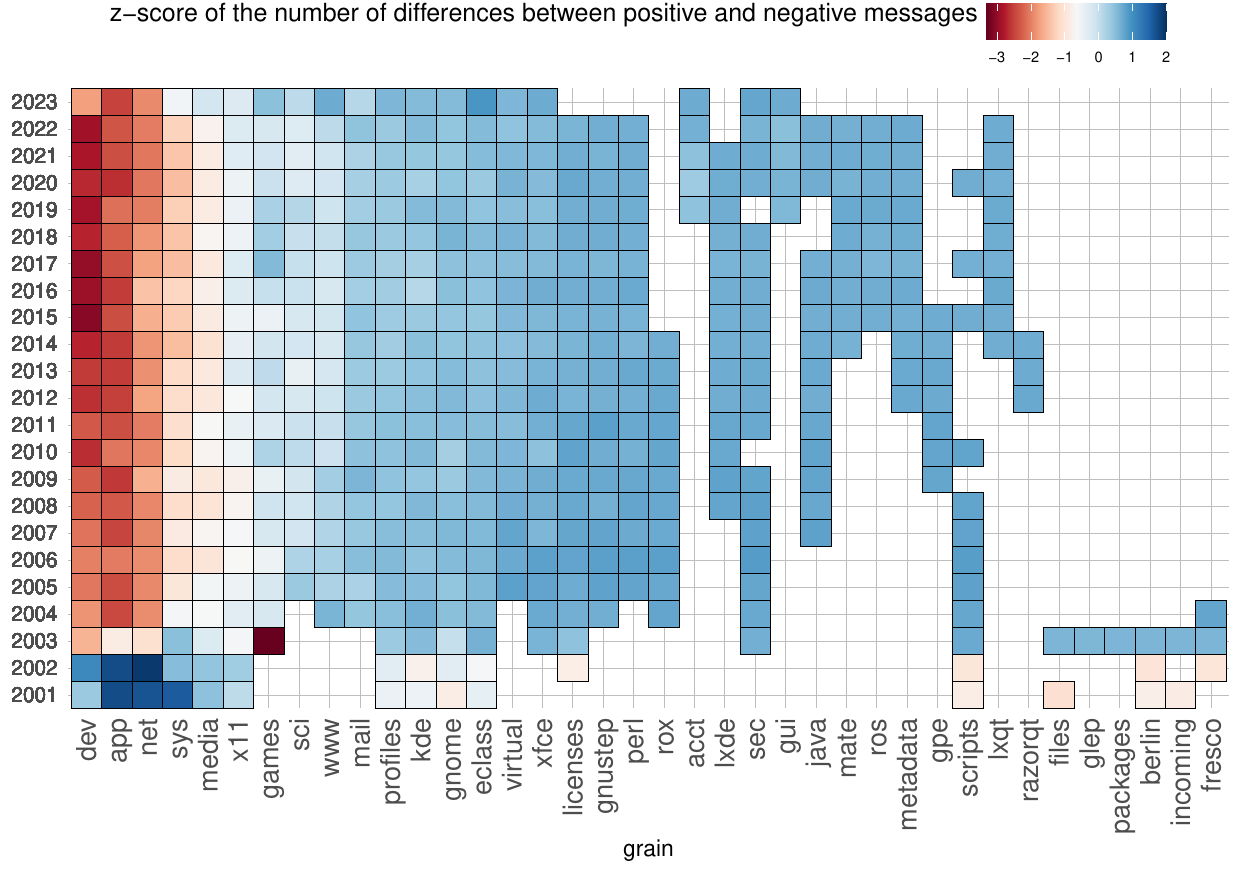}
    \caption{Heatmaps of number of negative and positive messages containing negative/positive sentences by grains with standard  normalisation (z-score) applied yearly}
    \label{fig_RQ2:_grains_normalised_yearly} 
\end{figure*}

To investigate RQ2, we visually inspected the results from the Gentoo components; for RQ3, we used a more structural statistical approach to determine how emotion in communication is associated with the relative number of activities during development.

As seen in Figure~\ref{fig_RQ2:_grains_normalised_yearly}, we found a total of 37 path grains during the Gentoo evolution: some of them existed throughout the observed period, while others were added or removed at various points in its history. In order to obtain an overall view of the negative and positive messages, we normalised the differences between positive and negative messages in all the grains yearly with a relative z-score scaling method because of the variance of the data~\cite{adeyemo2019effects}. The number of negative and positive messages varies for each path grain, on a scale of -3 and 2, between 2001 and 2023. The darker the red colour, the higher the number of negative messages sent in a grain; the darker the blue, the higher the number of positive messages in a path grain. 

The significant view from the heatmaps is that some grains clearly contained more negative messages than some other grains that had more positive messages. For example, grain \textit{‘dev’}, \textit{‘app’}, \textit{‘net’} which has a range of z-score less than -2 from 2003 to 2023, even though in the first two years, those grains were significantly positive (the z-scores range above 1). However, many more grains (more than 75\%) show more positivity (the z-scores were above 1) during all the period of time. 

\begin{sidewaysfigure*}
    \centering
    \includegraphics[width=1\linewidth]{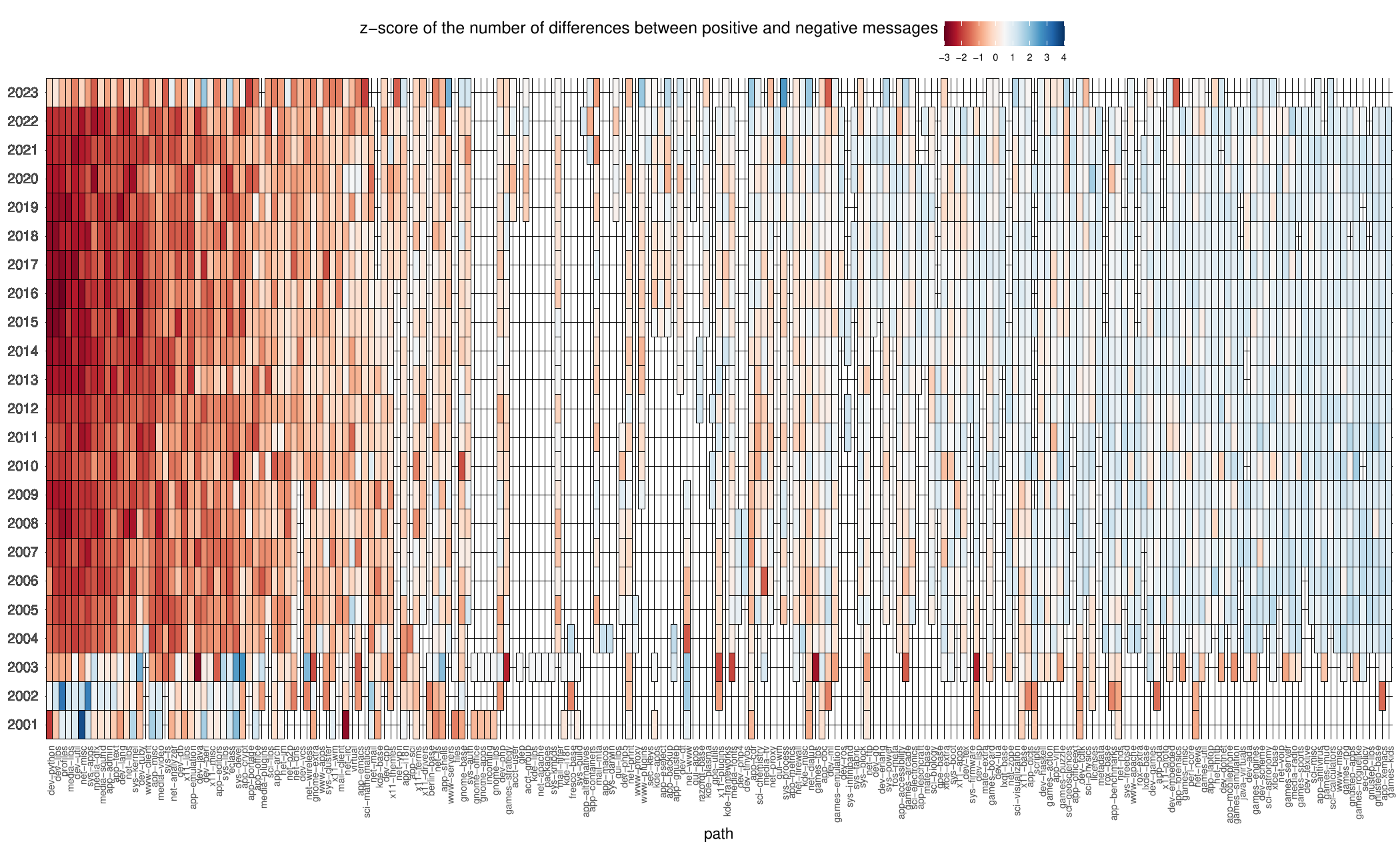}
    \caption{Heatmaps of number of negative and positive messages containing negative/positive sentences by paths with standard normalisation applied yearly} 
    \label{fig_RQ2:_fig_paths_normalised_yearly} 
\end{sidewaysfigure*}

The view on the grains performed above was further augmented by a similar analysis of the paths. This type of visualisation is more precise in terms of which components are affected by emotions: from the selection of paths displayed in Figure~\ref{fig_RQ2:_fig_paths_normalised_yearly}, for example: opting for the most top ten negative paths: ‘\textit{dev-python}’, `\textit{dev-libs}', `\textit{profiles}', `\textit{dev-utils}', `\textit{net-misc}', `\textit{sys-apps}', `\textit{media-libs}', `\textit{media-gfx}', and `\textit{media-sounds}', and `\textit{app-admin}', are all individual paths that suffer from negative messages. All of these paths have a z-score of less than -1. Nevertheless, looking at the most positive paths include `\textit{games-kids}', `\textit{app-xemacs}', `\textit{gnustep-base}', `\textit{sec-policy}', `\textit{games-roguelike}', `\textit{gnustep-apps}', `\textit{games-rpg}', `\textit{www-misc}', `\textit{sci-calculators}', and `\textit{games-mud}'. All of these paths have z-scores greater than +1.


\subsection{\textbf{Results: RQ3 -- The impact of the emotion on the Gentoo developers' activity}}

Using the distributions of z-score of the differences between DWPs and DWPs and the differences between positive negative messages across the Gentoo evolution in each grain and path, we evaluated their inter-quartile values (Q1 and Q3) to determine three categories (High, Medium and Low) for each attribute, as follows: ‘high’~(H) when values are $<$ Q3; ‘medium’~(M) when values are $\ge$ Q1 and $\le$Q3; ‘low’~(L) when values are $>$ Q1. 

\begin{table*}[ht]
\center
\caption{Classification using the Q1 and Q3 median values of the distribution on
sentiments (in grains).}
\label{tab_summary_median_grains}
\scalebox{0.80}{
    \begin{tabular}{|c|c|p{0.5cm}|p{0.5cm}|p{0.5cm}|p{0.5cm}|p{0.5cm}|p{0.5cm}|p{0.5cm}|p{0.5cm}|p{0.5cm}|p{0.5cm}|p{0.5cm}|p{0.5cm}|}
    \hline 
        \multicolumn{2}{|c|} {} & 
            \multicolumn{12}{|c|}{\# of grains containing normalised differences of sentiments} \\
    \cline{3-14}
        \multicolumn{2}{|c|} {} & 
            \multicolumn{3}{|c|}{2001-2003} & 
            \multicolumn{3}{|c|}{2005-2007} & 
            \multicolumn{3}{|c|}{2011-2013} & 
            \multicolumn{3}{|c|}{2019-2023} \\
        \cline{3-14}
        \multicolumn{2}{|c|}
            {} & $H^{M}_{p}$ & $M^{M}_{p}$ & $L^{M}_{p}$  & $H^{M}_{p}$ & $M^{M}_{p}$ & $L^{M}_{p}$ & $H^{M}_{p}$ & $M^{M}_{p}$ & $L^{M}_{p}$ & $H^{M}_{p}$ & $M^{M}_{p}$ & $L^{M}_{p}$\\ 
        \hline
            \# Grains containing & $H^{D}_{p}$ & 
            \cellcolor{HH}0 & \cellcolor{HM}0 & \cellcolor{HL}1 & 
            \cellcolor{HH}4 & \cellcolor{HM}0 & \cellcolor{HL}0 & 
            \cellcolor{HH}1 & \cellcolor{HM}3 & \cellcolor{HL}0 & 
            \cellcolor{HH}0 & \cellcolor{HM}1 & \cellcolor{HL}10 \\
             DWNs and DWP & $M^{D}_{p}$ & 
             \cellcolor{HM}0 & \cellcolor{MM}0 & \cellcolor{ML}1 & 
             \cellcolor{HM}5 & \cellcolor{MM}0 & \cellcolor{ML}0 & 
             \cellcolor{HM}1 & \cellcolor{MM}6 & \cellcolor{ML}1 & 
             \cellcolor{HM}0 & \cellcolor{MM}1 & \cellcolor{ML}7 \\
            (normalised differences)& $L^{D}_{p}$ & 
            \cellcolor{HL}0 & \cellcolor{ML}0 & \cellcolor{LL}12 & 
            \cellcolor{HL}9 & \cellcolor{ML}1 & \cellcolor{LL}2 & 
            \cellcolor{HL}3 & \cellcolor{ML}3 & \cellcolor{LL}3 & 
            \cellcolor{HL}0 & \cellcolor{ML}1 & \cellcolor{LL}3 \\
        \hline
        \multicolumn{2}{|c|} {Total Paths} & 
            \multicolumn{3}{|c|}{14} & 
            \multicolumn{3}{|c|}{21} & 
            \multicolumn{3}{|c|}{21} &
            \multicolumn{3}{|c|}{23}\\
    \hline
\end{tabular}}
\end{table*}

In our results, we normalised the differences between the counts of positive and negative messages, as well as the differences between the counts of positive and negative words. We implemented z-score or standard normalisation having the range values between negative and positive values. Hence, the more negative values refer to the high negativity contained in the packages or grains, and vice versa.

The years 2002, 2006 and 2012 were chosen as the peaks in the level of mailing list activity (as seen in Fig.~\ref{fig_RQ1:_histogram}). From those, three periods of years (2001-2003, 2005-2007 and 2011-2013), complemented with the latest available period (2019-2023), were chosen to evaluate each path and classify it into categories of `emotion within paths'. We statistically quantified the distributions of values of each attribute (the normalised differences between DWNs and DWPs and sentiments in the messages) and found two sets: 

\begin{itemize}
    \item $H^{D}_{g}$: $diff\_DWPs\_DWNs_{g} < -0.39$; $M^{D}_{g}$: $0.12 \ge diff\_DWPs\_DWNs_{g}\ge -0.39$; $L^{D}_{g}: 0.12 > diff\_DWPs\_DWNs_{g}$
    \item $H^{M}_{g}$: $EmoMes_{g} < -0.58$; $M^{M}_{g}$: $0.02 \ge EmoMes_{g} \ge -0.58$; $L^{M}_{g}: 0.02 > EmoMes_{g}$
\end{itemize}

where $H^{D}_{g}$ are high values of normalised differences between DWPs and DWNs in the grain $g$,  $M^{D}_{g}$ the medium ones, and $L^{D}_{g}$ the low ones; $diff\_DWPs\_DWNs_{g}$ is the median of the yearly values of normalised differences between DWPs and DWNs in the same grain $g$. Similarly, $H^{M}_{g}$ are high values of normalised differences between the counts of positive and negative messages in the grain $g$,  $M^{M}_{g}$ the medium ones, and $L^{M}_{g}$ the low ones; $EmoMes_{g}$ is the median value of normalised differences between the counts of positive and negative messages in the same grain $g$ during the analysed period. Table~\ref{tab_summary_median_grains} reports the results of our classification.

Table~\ref{tab_summary_median_grains} shows the evolution of grains in the categories: couples of high values ($H^{M}_{g} \cup H^{D}_{g}$) or medium-high values, ($H^{M}_{g} \cup M^{D}_{g}$, or $H^{D}_{g} \cup M^{M}_{g}$) signal those grains where communication issues have arose in a period, as a combination of developers' attributes and messages. 

The evolution of the communication within the Gentoo community is visible in the four chosen periods: some grains initially have a high number of $H^{M}_{g}$ and either $H^{D}_{g}$ or $M^{D}_{g}$ (9 out of 21 grains). However, the number of those ‘negative communication’ grains decreases in the second span (5 out of 21) and is significantly none in the last period. 

Complementing the analysis on the grains, a similar analysis was performed on the paths to obtain a more precise picture of where emotional communication occurs. Similarly to the above, we found the following H, M and L values:

\begin{itemize}
     \item{$H^{D}_{p}$: $diff\_DWPs\_DWNs_{p} < -0.49$; $M^{D}_{p}$: $0.5 \ge diff\_DWPs\_DWNs_{p} \ge -0.49$; $L^{D}_{p}: 0.5 > diff\_DWPs\_DWNs_{p}$}
     \item{$H^{M}_{p}$: $EmoMes_{p} < -0.76$; $M^{M}_{p}$: $0.68 \ge EmoMes_{p} \ge -0.76$; $L^{M}_{p}$: $0.68 > EmoMes_{p}$}
\end{itemize}

where $H^{D}_{p}$ are high values of normalised differences between DWPs and DWNs in the path $p$,  $M^{D}_{p}$ the medium ones, and $L^{D}_{p}$ the low ones; $DWNs_{p}$ are the yearly values of normalised differences between DWPs and DWNs in the same path $g$. Similarly, $H^{M}_{p}$ are high values of normalised differences between the counts of positive and negative messages the path $p$,  $M^{M}_{p}$ the medium ones, and $L^{M}_{p}$ the low ones; $NegMes_{p}$ are the yearly values of normalised differences between the counts of positive and negative messages in the same path $p$. Table~\ref{tab_summary_median_paths} reports our classification at the path level. 

From table~\ref{tab_summary_median_paths}, only seven paths initially suffered from a pattern of ‘negative communication’; subsequently, 73 out of 144 paths (around 51\%) suffered from the pattern during the second period (\textit{i.e.}, $H^{M}_{p} \cup H^{D}_{p}$, $H^{M}_{p} \cup M^{D}_{p}$ or $H^{D}_{p} \cup M^{M}_{p}$). This set was reduced to 44 paths (36 out of 144 or 25\%) between 2011 and 2013 and further reduced to 28 (out of 148, around 19\%) between 2019 and 2023.  

\begin{table*}
\center
\caption{Classification using the Q1 and Q3 median values of the distribution on
sentiments (in paths).}
\label{tab_summary_median_paths}
\scalebox{0.80}{
    \begin{tabular}{|c|c|p{0.5cm}|p{0.5cm}|p{0.5cm}|p{0.5cm}|p{0.5cm}|p{0.5cm}|p{0.5cm}|p{0.5cm}|p{0.5cm}|p{0.5cm}|p{0.5cm}|p{0.5cm}|}
    \hline 
        \multicolumn{2}{|c|} {} & 
            \multicolumn{12}{|c|}{\# of paths containing normalised differences of sentiments} \\
    \cline{3-14}
        \multicolumn{2}{|c|} {} & 
            \multicolumn{3}{|c|}{2001-2003} & 
            \multicolumn{3}{|c|}{2005-2007} & 
            \multicolumn{3}{|c|}{2011-2013} & 
            \multicolumn{3}{|c|}{2019-2023} \\
        \cline{3-14}
        \multicolumn{2}{|c|}
            {} & $H^{M}_{p}$ & $M^{M}_{p}$ & $L^{M}_{p}$  & $H^{M}_{p}$ & $M^{M}_{p}$ & $L^{M}_{p}$ & $H^{M}_{p}$ & $M^{M}_{p}$ & $L^{M}_{p}$ & $H^{M}_{p}$ & $M^{M}_{p}$ & $L^{M}_{p}$\\ 
        \hline
            \# Paths containing & $H^{D}_{p}$ & 
            \cellcolor{HH}0 & \cellcolor{HM}7 & \cellcolor{HL}37 & \cellcolor{HH}21 & \cellcolor{HM}9 & \cellcolor{HL}1 & \cellcolor{HH}2 & \cellcolor{HM}17 & \cellcolor{HL}3 & \cellcolor{HH}0 & \cellcolor{HM}28 & \cellcolor{HL}5 \\
             DWNs and DWP & $M^{D}_{p}$ & \cellcolor{HM}0 & \cellcolor{MM}9 & \cellcolor{ML}23 & \cellcolor{HM}43 & \cellcolor{MM}11 & \cellcolor{ML}0 & \cellcolor{HM}17 & \cellcolor{MM}69 & \cellcolor{ML}3
             & \cellcolor{HM}0 & \cellcolor{MM}55 & \cellcolor{ML}35\\
            (normalised differences)& $L^{D}_{p}$ & \cellcolor{HL}0 & \cellcolor{ML}2& \cellcolor{LL}13 & \cellcolor{HL}52 & \cellcolor{ML}7 & \cellcolor{LL}0 & \cellcolor{HL}10 & \cellcolor{ML}21 & \cellcolor{LL}2
            & \cellcolor{HL}0 & \cellcolor{ML}18 & \cellcolor{LL}7\\
        \hline
        \multicolumn{2}{|c|} {Total Paths} & 
            \multicolumn{3}{|c|}{91} & 
            \multicolumn{3}{|c|}{144} & 
            \multicolumn{3}{|c|}{144} &
            \multicolumn{3}{|c|}{148}\\
    \hline
\end{tabular}}
\end{table*}

\begin{figure*}[h]
    \centering
    \includegraphics[width=.9\linewidth]{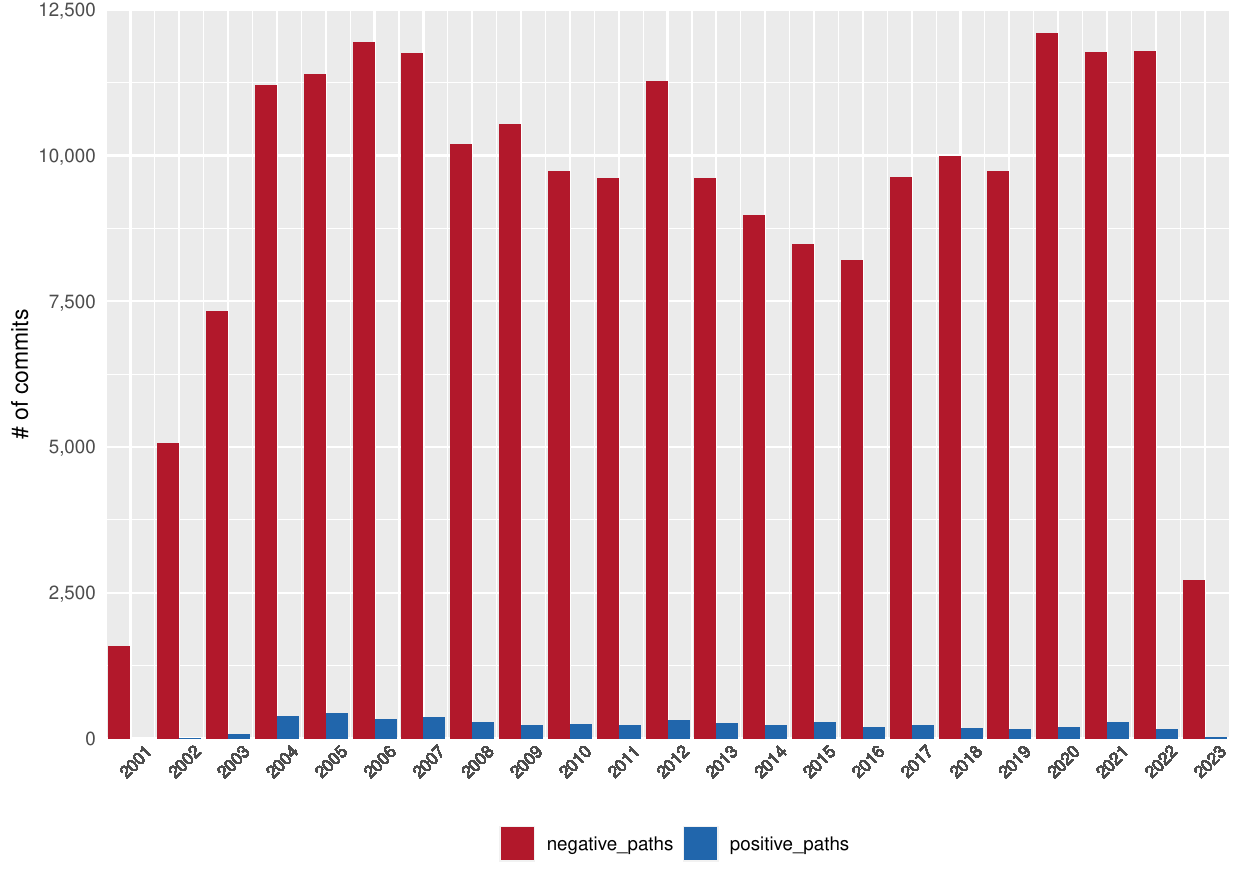}
    \caption{bar graphs of the number of commits of top ten negative paths and top ten (red-color bars) positive paths (blue-color bars)} 
    \label{fig_RQ3:_fig_paths_commits} 
\end{figure*}

Investigating deeper the impact of the emotions shown by the developers on their activity, we observed that the commits made on the packages suffered high negativity and high positivity. We sampled the top ten negative paths and top ten positive paths shown by Figure~\ref{fig_RQ3:_fig_paths_commits}. We statistically calculated the Wilcoxon on two groups consisting of the number of commits of developers working on the top ten negative paths and top ten positive paths that we have mentioned in the section~\ref{sec:_results_RQ2a_extent_emotions_on_components}. We discovered the p-value = 2.2e-16 with the confidence interval 95\%. Hence, we rejected the null hypothesis: "The number of commits done in the most negative paths and in the most positive paths are equal". In other words, we concluded that more negative communication was seen in the higher commit activity.

\section{Discussion and Implications}
\label{sec:_discussion_and_implications}

\subsection{Evolution of the emotions in Gentoo project}
Our study extends the findings of a previous analysis of the Gentoo community \cite{garcia2013role}, by adding evidence on the last 11 years of its development. As found in the earlier study, Gentoo developers showed more negative emotions during some pivotal events (2006 and 2012), when the procedures in the community were modified and restructured. Also, the event of one (or more) of the main contributors leaving the community triggered high spikes in negativity among developers. This has been documented in the first decade of Gentoo's evolution when key contributors left the community~\cite{garcia2013role, GentooBugWranglers}. Additionally, the drop in the number of messages in 2008 was triggered by the mass resignation from the Board members of the Gentoo~\cite{}.

In our analysis, we complemented the former study by showing that, over two decades, both negative and positive sentiment can be clearly detected in the communication between developers, and along the basic components of the Gentoo project; differently from the former paper, our results have also shown a major change in the sentiment in the mailing list, as the level of both negative and positive communication has been drastically reduced. The mailing list used in our study is one of the communication means employed by Gentoo developers to discuss technical issues during development\cite{GentooMailingList}. We informally reached out to some of the developers contributing to the Gentoo community. Their reports mentioned some possible reasons why the messages on the mailing list declined. Most of the causes mentioned are the shifting to another dynamic platform, such as Github, and more real-time, chat-based platforms, such as IRC and Matrix. Additionally, the disengagement of the project among the developers and also the decrease of vocal participants from the long-term contributors may also impact the communication frequency in the mailing list.

\subsection{Grains and paths} The higher negative communication in certain \textit{grains} and \textit{paths} is a proxy of components where many developers work in parallel and a signal of those components being essential parts of the project, as corrective maintenance of the packages~\cite{licorish2018exploring} and bug-related tasks are likely to contain negative emotions~\cite{Huq2020, graziotin2018happens}. Relating to work in parallel, one of the developers that we have reached out to mentioned the shortage of development resources: \textit{"each of main developers usually has to juggle a lot of different at the same time: toolchain, packages, testing, new internal tooling, etc. ... it's impossible to keep up the pace with large projects".}

Several grains, including \textit{`dev'}, \textit{`app'}, and \textit{`net'} were highly impacted by negativity compared to other grains during all the years. Inside these grains, we could clearly isolate specific paths, namely \textit{`dev-python'}, \textit{`dev-libs'}, \textit{`dev-utils'}, \textit{`net-misc'}, \textit{`app-admin'} that have been highly impacted by negative communication. The importance of the Gentoo project of those grains and paths (and the packages contained) could be one of the causes of the negativity expressed by developers. For instance, the ‘\textit{dev-python}’ path shows high negative communication: as a component, it provides essential packages related to Python modules, extensions and bindings, tools and utilities useful for development in Python~\cite{GentooLinuxPackages}. Additionally, Python, as an open-source and cross-platform language, may have great demand from users~\cite{IsPythoninHighDemand}. Hence making it to having high-maintenance among the developers.  

Moreover, regarding the aforementioned paths, one developer mentioned that those paths were part of the core system utilities whose configurations have a significant impact on a Gentoo system as a whole, hence having different arguments for different approaches to managing such packages. Additionally, a path such as \textit{`eclass'} containing common code modules used by many different packages may have arguments over the development of particular modules in that path. Besides the disagreements issue, the
developers also mentioned the difficulty of maintaining the critical core component. One existing issue that caused the negative spike between 2006 and 2007 was that one developer broke \textit{`glibc'} (one of the critical core components).

\subsection{Implications for practitioners}
\textbf{\textit{The importance of fostering a positive and constructive communication culture within the open-source software development community.}} The increase in negative messages outweighing positive ones found in our study suggests a shift in the tone of communication within the software development community. The practitioners should be mindful of the impact of negative communication on team dynamics, collaboration, and overall project success. The leading or maintainers may implement methods aimed at cultivating a communication environment that is both more optimistic and productive. For example, implementing the code of conduct regulating acceptable or unacceptable communication along with the consequences for those violating the rules. In addition, creating a more conducive and welcome-respectful workplace may encourage new developers to join~\cite{tourani2017code,trinkenreich2022women,bayati2019effect}.

\textbf{\textit{Understanding the evolution of the negative emotions}} Examining how negative emotions have changed within the Gentoo project over 23 years can aid practitioners in recognising patterns of emotional evolution. The trends may provide insights to implement ad-hoc strategies (for instance, to mitigate controversies during code reviews~\cite{el2019empirical}) to assess and prevent negative emotions in the future. For example, the practitioners may develop clear and comprehensive guidelines for conducting code reviews, outlining expected behaviour, communication norms, and conflict resolution procedures. These guidelines provide a framework for productive and respectful interactions during code reviews, helping to prevent misunderstanding and reduce negative emotions.

\textbf{\textit{Observation on developer behaviour}}. The observation that the most active developers tend to use fewer negative messages than the top negative senders highlights potential differences in communication styles and their impact on project dynamics. Practitioners can analyse these patterns to understand better how individual developer behaviour influences the overall tone and effectiveness of communication within the community. For example, profiling the behaviour of individual developers based on their communication patterns, activity levels, and contributions to the project may help the leading developers to inform talent management, team formation, and resource allocation decisions within the community.

\textbf{\textit{Investigating on the granular level of the components.}} The declining pattern observed in grains and paths associated with negative communication across various time intervals underscores the importance of monitoring patterns at a granular level. Practitioners can leverage this information to identify specific components within the project where negative communication is prevalent and implement targeted interventions to address underlying issues. For example, regular code reviews should be implemented specifically focused on certain critical components that are prone to issues and bugs.

\subsection{Implications for researchers}
\textbf{\textit{A comprehensive and multi-faceted investigation of written communication over two decades.}}
Our dataset contains data from a large open-source community spanning 23 years. This dataset is more than a starting point to investigate the developer's behaviours: it provides a long-term, multi-faceted view into written communication containing stress, sadness, depression, and anger. Further research can explore the relationship between developer behaviour, communication styles, and project outcomes. The researchers may investigate why certain developers use more negative or positive messages than others and how these communication patterns impact collaboration, productivity, and code quality. Hence, these can provide valuable insights into effective teamwork and software development practices.

\textbf{\textit{Qualitative research for deeper insights.}} 
While our quantitative analysis provides valuable insights, researchers can complement these findings with qualitative methods such as interviews and surveys. This can further enrich the understanding of the underlying motivations, perceptions, and experiences of developers within the open-source software community, providing richer insights into communication dynamics. 

\textbf{\textit{Comparative studies and generalisability.}} Our methodology may be expanded to different open-source projects or software development communities. Researchers can conduct comparative studies comparing communication patterns, developer behaviours, and project outcomes across diverse contexts. This can help identify common trends and context-specific factors influencing dynamics. 

\textbf{\textit{Path Analysis and commit activity.}} The declining pattern observed in grains and paths associated with negative communication suggests a potential correlation between communication dynamics and software development activity. Researchers can explore the link between negative communication paths and heightened commit activity, examining how communication patterns influence developer engagement, productivity, and code quality. For example, the researcher could conduct a longitudinal study within a software development team working on a large-scale project and collect communication data over several months.

\section{Threats to validity}
\label{sec:_threats to validity}
\subsection{Construct Validity}
We realise that we cannot use an absolute number of positive and negative messages to identify a DWP and a DWN, respectively, since each developer has a different engagement length in the Gentoo project. Therefore, we have performed the normalisation by dividing the number of positive or negative messages by the total number of messages written by each developer. Moreover, the definition of the path grain derived from the similar first names of the paths may generalise the packages without considering the real themes behind the packages' names. We believe that defining them into one-grain names may give the main idea of what the packages are about. Additionally, the response from the Gentoo developers also supports this definition, for example `\textit{x11}' is related to the display-manager renewal. Hence, the path grain definition poses little threat to the construct validity.

\subsection{Conclusion Validity}
One limitation of our work is that we consider the absolute number of messages written instead of the ratio between the `number of messages' and `number of negative messages'. Our figures are constructed based on this. Applying the proportion might yield different outputs of the peak years for negative messages. Furthermore, our sentiment labels (at the sentence level) entirely rely on the Sentistrength-SE tool. This may generalise the sentence's meaning without thoroughly examining the context of the sentence itself or the whole message, which may have a different meaning. Therefore, we conducted the manual annotations on the sampling of sentence collection to mitigate this threat. Furthermore, mailing lists have a range of topics and ambiguity in the discussion; therefore, we did some sampling by searching several terms related to paths (e.g., `\textit{dev}', `\textit{app}', `\textit{net}', `\textit{media}', and, `\textit{sys}'). We found roughly 3-8\% messages mentioning the subset of path names of the total messages in the mailing list. 

\subsection{Internal and External Validity}
Our analysis only investigates the messages taken from the mailing list. The community also employs other means of communication, such as Bugzilla reports, IRC, and discussion forums. In addition, we only joined the datasets between the same dates, as we also joined the tables from the mailing list and the commits datasets. We did not consider that some messages sent on different dates (prior to or after the considered interval) may be related to some commits from different dates. 

We only relied upon our analysis from one open-source community, suggesting that our findings may not be generalisable. Therefore, replications to the different open-source communities are suggested. In order to assess the verifiability of our methods and results and to support the independent replication of our study, we provide a link\footnote{https://shorturl.at/DJ023} with free access and anonymity.

\section{Conclusion and Future Work}
\label{sec:_conclusion}
In this work, we have investigated the extent of negative emotions shown in the messages of the Gentoo mailing list during the evolution of this OSS community. We have also characterised how negative emotions impacted the projects' file paths (as components). 

We found that our approach can precisely identify which components have been structurally affected by negative emotions. We also found that complementing an earlier study, the level of negative emotions expressed in the mailing lists decreased in the last 23 years and that this reduction occurred (in a more or less pronounced way) along Gentoo's development paths.

The approach we have presented here can be applied to any software project where written communication and development logs are available. Our work has the potential to highlight which components have suffered (or still suffer) from negative communication and isolate the components that are affected by both negative language and developers sending negative messages.

We believe our work to be one step in the right direction: having observed a correlation between negative sentiment and productivity, the next logical step will be to characterise each component in terms of its \textit{functionality} and to investigate whether specific functionalities or their complexity, are more strongly linked to negative communication. For this, we also planned to extend our datasets with bug reports and instant messaging logs.

Finally, we plan to expand our analysis qualitatively, delving into the messages' content and contacting the developers with a questionnaire. Also, since the Sentistrength-SE is limited to binary classifications, we will consider other multi-class tools to detect stress, sadness, depression, and anger. We believe that taking more emotions into consideration in classifying developers' emotional states may give a more comprehensive representation of individual emotions.

\bmsection*{Author contributions}
All the authors contributed equally.

\bmsection*{Acknowledgments}
The authors would like to express their gratitude to V. Soancatl Aguilar, PhD from the Department of Computer and Information Technology, University of Groningen, for his invaluable insights and suggestions on visualisation in our work. Tien Rahayu Tulili gratefully acknowledges scholarship funding for this research from the Indonesia Endowment Fund for Education(LPDP), Ministry of Finance of the Republic of Indonesia, Ref. Number S-2566/LPDP.4/2021.

\bmsection*{Conflict of interest}
The authors declare no potential conflict of interest.

\bibliography{main}

\begin{thebibliography}{10}
\providecommand \doibase [0]{http://dx.doi.org/}%

\bibitem{narduzzo2005role}
Narduzzo A, Rossi A. The role of modularity in free/open source software development. In: , , Igi Global,  2005\string:84--102.

\bibitem{langlois2008hackers}
Langlois RN, Garzarelli G. Of hackers and hairdressers: Modularity and the organizational economics of open-source collaboration. {\it Industry and Innovation.} 2008\string;15(2)\string:125--143.

\bibitem{mockus2000case}
Mockus A, Fielding RT, Herbsleb J. A case study of open source software development: the Apache server. In: Association for Computing Machinery.  2000\string:263--272.

\bibitem{murgia2018exploratory}
Murgia A, Ortu M, Tourani P, Adams B, Demeyer S. An exploratory qualitative and quantitative analysis of emotions in issue report comments of open source systems. {\it Empirical Software Engineering.} 2018\string;23\string:521--564.

\bibitem{murgia2014developers}
Murgia A, Tourani P, Adams B, Ortu M. Do developers feel emotions? an exploratory analysis of emotions in software artifacts. In: Association for Computing Machinery.  2014\string:262--271.

\bibitem{Sharp2015ClosingADoor}
{Sage Sharp} . Closing a door. Web page;  2015.
\newblock \url{https://sage.thesharps.us/2015/10/05/closing-a-door/}, Last accessed on 09/2023.

\bibitem{Ranzhin2019Iruin}
{Philipp Ranzhin} . I ruin developer's lives with my code reviews and I'm sorry. Web page;  2015.
\newblock \url{https://habr.com/en/articles/440736/}, Last accessed on 09/2019.

\bibitem{gachechiladze2017anger}
Gachechiladze D, Lanubile F, Novielli N, Serebrenik A. Anger and its direction in collaborative software development. In: IEEE.  2017\string:11--14.

\bibitem{Huq2020}
Huq SF, Sadiq AZ, Sakib K. Is Developer Sentiment Related to Software Bugs: An Exploratory Study on GitHub Commits. In: the Institute of Electrical and Electronics Engineers.  2020\string:527-531.

\bibitem{robinson2016developer}
Robinson WN, Deng T, Qi Z. Developer behavior and sentiment from data mining open source repositories. In: IEEE.  2016\string:3729--3738.

\bibitem{madampe2020towards}
Madampe K, Hoda R, Singh P. Towards understanding emotional response to requirements changes in agile teams. In: Association for Computing Machinery/IEEE.  2020\string:37--40.

\bibitem{garcia2013role}
Garcia D, Zanetti MS, Schweitzer F. The role of emotions in contributors activity: A case study on the Gentoo community. In: IEEE.  2013\string:410--417.

\bibitem{Anonymous2014}
{Anonymomus Author} . Leaving toxic OS communities. Web page;  2014.
\newblock Last accessed 09/2019.

\bibitem{raman2020stress}
Raman N, Cao M, Tsvetkov Y, K{\"a}stner C, Vasilescu B. Stress and burnout in open source: Toward finding, understanding, and mitigating unhealthy interactions. In: the Institute of Electrical and Electronics Engineers.  2020\string:57--60.

\bibitem{StackOverflowBlog}
Hanlon J. Stack Overflow Isn’t Very Welcoming. It’s Time for That to Change.. Web page;  2018.

\bibitem{zanetti2013rise}
Zanetti MS, Scholtes I, Tessone CJ, Schweitzer F. The rise and fall of a central contributor: Dynamics of social organization and performance in the Gentoo community. In: IEEE.  2013\string:49--56.

\bibitem{wu2023social}
Wu J, Huang X, Wang B. Social-technical network effects in open source software communities: understanding the impacts of dependency networks on project success. {\it Information Technology \& People.} 2023\string;36(2)\string:895--915.

\bibitem{tee2019modular}
Tee R, Davies A, Whyte J. Modular designs and integrating practices: Managing collaboration through coordination and cooperation. {\it Research policy.} 2019\string;48(1)\string:51--61.

\bibitem{graziotin2018happens}
Graziotin D, Fagerholm F, Wang X, Abrahamsson P. What happens when software developers are (un)happy. {\it Jnl of Systems and Software.} 2018\string;140\string:32--47.

\bibitem{ortu2016arsonists}
Ortu M, Destefanis G, Counsell S, Swift S, Tonelli R, Marchesi M. Arsonists or firefighters? Affectiveness in agile software development. In: Springer International Publishing.  2016\string:144--155.

\bibitem{wang2019emotions}
Wang Y. Emotions extracted from text vs. true emotions--an empirical evaluation in SE context. In: the Institute of Electrical and Electronics Enginners.  2019\string:230--242.

\bibitem{kim2018human}
Kim Y, Kim J. Human-like emotion recognition: Multi-label learning from noisy labeled audio-visual expressive speech. In: IEEE.  2018\string:5104--5108.

\bibitem{islam2018deva}
Islam MR, Zibran MF. DEVA: sensing emotions in the valence arousal space in software engineering text. In: Association for Computing Machinery.  2018\string:1536--1543.

\bibitem{islam2019marvalous}
Islam MR, Ahmmed MK, Zibran MF. MarValous: Machine learning based detection of emotions in the valence-arousal space in software engineering text. In: Association for Computing Machinery.  2019\string:1786--1793.

\bibitem{das2023aligning}
Das M. Aligning Emotions, Thoughts, and Feelings to Build a High-Performing Team. {\it Global journal of Business and Integral Security.} 2023.

\bibitem{rafaeli2009sensemaking}
Rafaeli A, Ravid S, Cheshin A. Sensemaking in virtual teams: The impact of emotions and support tools on team mental models and team performance. {\it International review of industrial and organizational psychology.} 2009\string:151--181.

\bibitem{kazemitabar2024examining}
Kazemitabar M, Lajoie SP, Doleck T. Examining the Relationship Between Socially-Shared Emotion Regulation and Building Team Coordination Mechanisms During a Hackathon. {\it Education and Information Technologies.} 2024\string;29(5)\string:6241--6272.

\bibitem{werder2018evolution}
Werder K. The evolution of emotional displays in open source software development teams: an individual growth curve analysis. In: Association for Computing Machinery.  2018\string:1--6.

\bibitem{von2003community}
Von~Krogh G, Spaeth S, Lakhani KR. Community, joining, and specialization in open source software innovation: a case study. {\it Research policy.} 2003\string;32(7)\string:1217--1241.

\bibitem{islam2018sentistrength}
Islam MR, Zibran MF. SentiStrength-SE: Exploiting domain specificity for improved sentiment analysis in software engineering text. {\it Journal of Systems and Software.} 2018\string;145\string:125--146.

\bibitem{calefato2018sentiment}
Calefato F, Lanubile F, Maiorano F, Novielli N. Sentiment polarity detection for software development. In: Association for Computing Machinery.  2018\string:128--128.

\bibitem{Pletea2014}
Pletea D, Vasilescu B, Serebrenik A. Security and emotion: sentiment analysis of security discussions on github. In: Association for Computing Machinery.  2014\string:348--351.

\bibitem{chen2021emoji}
Chen Z, Cao Y, Yao H, et al. Emoji-powered sentiment and emotion detection from software developers’ communication data. {\it ACM Transactions on Software Engineering and Methodology (TOSEM).} 2021\string;30(2)\string:1--48.

\bibitem{el2019empirical}
El~Asri I, Kerzazi N, Uddin G, Khomh F, Idrissi MJ. An empirical study of sentiments in code reviews. {\it Inf. and Soft. Technology (IST).} 2019\string;114\string:37--54.

\bibitem{girardi2021emotions}
Girardi D, Lanubile F, Novielli N, Serebrenik A. Emotions and perceived productivity of software developers at the workplace. {\it IEEE Transactions on Software Engineering.} 2021\string;48(9)\string:3326--3341.

\bibitem{richter2010words}
Richter M, Eck J, Straube T, Miltner WH, Weiss T. Do words hurt? Brain activation during the processing of pain-related words. {\it Pain.} 2010\string;148(2)\string:198--205.

\bibitem{adeyemo2019effects}
Adeyemo A, Wimmer H, Powell LM. Effects of normalization techniques on logistic regression in data science. {\it Journal of Information Systems Applied Research.} 2019\string;12(2)\string:37.

\bibitem{GentooBugWranglers}
{Gentoo Authors} . Bug Wranglers. Web page;  2001-2024.
\newblock \url{https://wiki.gentoo.org/wiki/Project:Bug-wranglers}.

\bibitem{GentooMailingList}
{Gentoo} . Gentoo Mailing List Archive. Web page;  2001-2020.
\newblock \url{https://archives.gentoo.org/, Last accessed on 04/2024}.

\bibitem{licorish2018exploring}
Licorish SA, MacDonell SG. Exploring the links between software development task type, team attitudes and task completion performance: Insights from the Jazz repository. {\it Information and software technology.} 2018\string;97\string:10--25.

\bibitem{GentooLinuxPackages}
{Gentoo Authors} . Gentoo Linux Packages. Web page;  2001-2023.
\newblock \url{https://packages.gentoo.org/categories/dev-python}, Last accessed on 05/2023.

\bibitem{IsPythoninHighDemand}
Petra . Is Python in high demand?. Web page;  2023.

\bibitem{tourani2017code}
Tourani P, Adams B, Serebrenik A. Code of conduct in open source projects. In: IEEE.  2017\string:24--33.

\bibitem{trinkenreich2022women}
Trinkenreich B, Wiese I, Sarma A, Gerosa M, Steinmacher I. Women’s participation in open source software: A survey of the literature. {\it ACM Transactions on Software Engineering and Methodology (TOSEM).} 2022\string;31(4)\string:1--37.

\bibitem{bayati2019effect}
Bayati S. Effect of Newcomers' Supportive Strategies on Open Source Projects Socio-Technical Activities. In: IEEE.  2019\string:49--50.

\bibitem{metallinou2008audio}
Metallinou A, Lee S, Narayanan S. Audio-visual emotion recognition using gaussian mixture models for face and voice. In: IEEE.  2008\string:250--257.

\bibitem{li2009emotion}
Li M, Lu BL. Emotion classification based on gamma-band EEG. In: the Institute of Electrical and Electronics Enginners.  2009\string:1223--1226.

\bibitem{cheriyan2021towards}
Cheriyan J, Savarimuthu BTR, Cranefield S. Towards offensive language detection and reduction in four Software Engineering communities. In: , , ,  2021\string:254--259.

\bibitem{ferreira2021shut}
Ferreira I, Cheng J, Adams B. The "shut the f** k up" phenomenon: Characterizing incivility in open source code review discussions. {\it Proceedings of the ACM on Human-Computer Interaction.} 2021\string;5(CSCW2)\string:1--35.

\bibitem{qiu2022detecting}
Qiu HS, Vasilescu B, K{\"a}stner C, Egelman C, Jaspan C, Murphy-Hill E. Detecting interpersonal conflict in issues and code review: cross pollinating open-and closed-source approaches. In: Association for Computing Machinery/IEEE.  2022\string:41--55.

\bibitem{lanubile2010collaboration}
Lanubile F, Ebert C, Prikladnicki R, Vizca{\'\i}no A. Collaboration tools for global software engineering. {\it IEEE software.} 2010\string;27(2)\string:52.

\bibitem{nasukawa2003sentiment}
Nasukawa T, Yi J. Sentiment analysis: Capturing favorability using natural language processing. In: Association for Machinery.  2003\string:70--77.

\bibitem{wiebe2000learning}
Wiebe J, others . Learning subjective adjectives from corpora. {\it Aaai/iaai.} 2000\string;20(0)\string:0.

\bibitem{tong2001operational}
Tong RM. An operational system for detecting and tracking opinions in on-line discussion. In: . 1. Association for Computing Machinery.  2001.

\bibitem{morinaga2002mining}
Morinaga S, Yamanishi K, Tateishi K, Fukushima T. Mining product reputations on the web. In: Association for Computing Machinery.  2002\string:341--349.

\bibitem{ng2002improving}
Ng V, Cardie C. Improving machine learning approaches to coreference resolution. In: Association for Computational Linguistics.  2002\string:104--111.

\bibitem{tan2016introduction}
Tan PN, Steinbach M, Kumar V. {\it Introduction to data mining}.
\newblock Pearson Education India, 2016.

\bibitem{han2014speech}
Han K, Yu D, Tashev I. Speech emotion recognition using deep neural network and extreme learning machine. In: Microsoft.  2014.

\bibitem{guzman2014sentiment}
Guzman E, Az{\'o}car D, Li Y. Sentiment analysis of commit comments in GitHub: an empirical study. In: Association for Computing Machinery.  2014\string:352--355.

\bibitem{sarker2020benchmark}
Sarker J, Turzo AK, Bosu A. A benchmark study of the contemporary toxicity detectors on software engineering interactions. In: IEEE.  2020\string:218--227.

\bibitem{steinmacher2015social}
Steinmacher I, Conte T, Gerosa MA, Redmiles D. Social barriers faced by newcomers placing their first contribution in open source software projects. In: Association for Computing Machinery.  2015\string:1379--1392.

\bibitem{Gentoo}
{Gentoo Authors} . Project:RelEng. Web page;  2023.
\newblock \url{https://wiki.gentoo.org/wiki/Project:RelEng}, Last accessed on April, 2023.

\bibitem{DistroWatch}
{Atea Ataroa Limited} . DistroWatch.com. Web page;  2023.
\newblock \url{https://distrowatch.com/index.php?distribution=gentoo}, Last accessed on 04/2023.

\bibitem{giuffrida2013empirical}
Giuffrida R, Dittrich Y. Empirical studies on the use of social software in global software development--A systematic mapping study. {\it Information and Software Technology.} 2013\string;55(7)\string:1143--1164.

\bibitem{pennebaker1999linguistic}
Pennebaker JW, King LA. Linguistic styles: language use as an individual difference.. {\it Journal of personality and social psychology.} 1999\string;77(6)\string:1296.

\bibitem{van2021promises}
Mil vFC, Rastogi A, Zaidman A. Promises and Perils of Inferring Personality on GitHub. In: Association for Computing Machinery/IEEE.  2021\string:1--11.

\bibitem{Slashdog}
{SlashdogMedia} . Slashdog.org. Web page;  2018.
\newblock \url{https://developers.slashdot.org/story/04/04/26/2259211/daniel-robbins-resigns-as-chief-gentoo-architect}, Last accessed on 05/2023.

\bibitem{GentooLinuxPackagesDev}
{Gentoo Authors} . Gentoo Linux Packages. Web page;  2001-2023.
\newblock \url{https://packages.gentoo.org/categories/dev-util}, Last accessed on 05/2023.

\bibitem{rastogi2016personality}
Rastogi A, Nagappan N. On the personality traits of GitHub contributors. In: IEEE.  2016\string:77--86.

\bibitem{leonardo2023}
{Leonardo Montini} . I QUIT from a TOXIC Workplace for Developers. Web page;  2023.
\newblock https://dev.to/this-is-learning/i-quit-from-a-toxic-software-development-workplace-4g3a.

\bibitem{herbsleb2006collaboration}
Herbsleb J, Roberts J. Collaboration in software engineering projects: A theory of coordination. In: Association for Information Systems.  2006.

\bibitem{herbsleb2007global}
Herbsleb JD. Global software engineering: The future of socio-technical coordination. In: IEEE.  2007\string:188--198.

\bibitem{chung2023impact}
Chung M, Sharma L, Malhotra MK. Impact of Modularity Design on Mobile App Launch Success. {\it Manufacturing \& Service Operations Management.} 2023\string;25(2)\string:756--774.

\bibitem{ralph2020empirical}
Ralph P, Ali Nb, Baltes S, et al. Empirical standards for software engineering research. {\it arXiv preprint arXiv:2010.03525.} 2020.

\end{thebibliography}

\bmsection*{Supporting information}

Additional supporting information may be found in the
online version of the article at the publisher’s website.

\appendix
\bmsection{The figures\label{app1}}
\begin{sidewaysfigure*}
    \centering
    \includegraphics[width=0.85\linewidth]{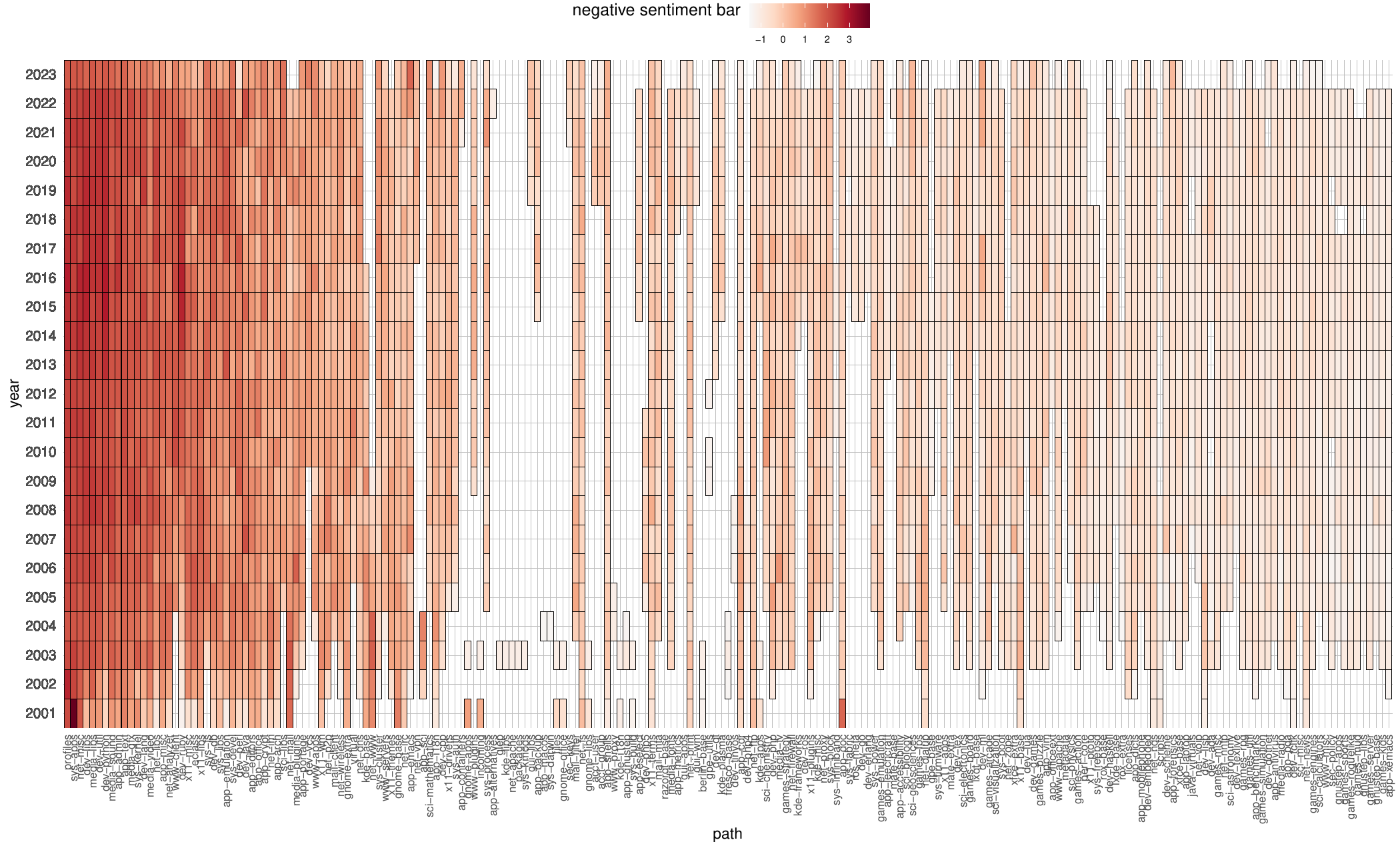}
    \caption{Heatmaps of number of negative  messages containing negative sentences by paths with standard normalisation applied yearly} 
\end{sidewaysfigure*}

\begin{sidewaysfigure*}
    \centering
    \includegraphics[width=1\linewidth]{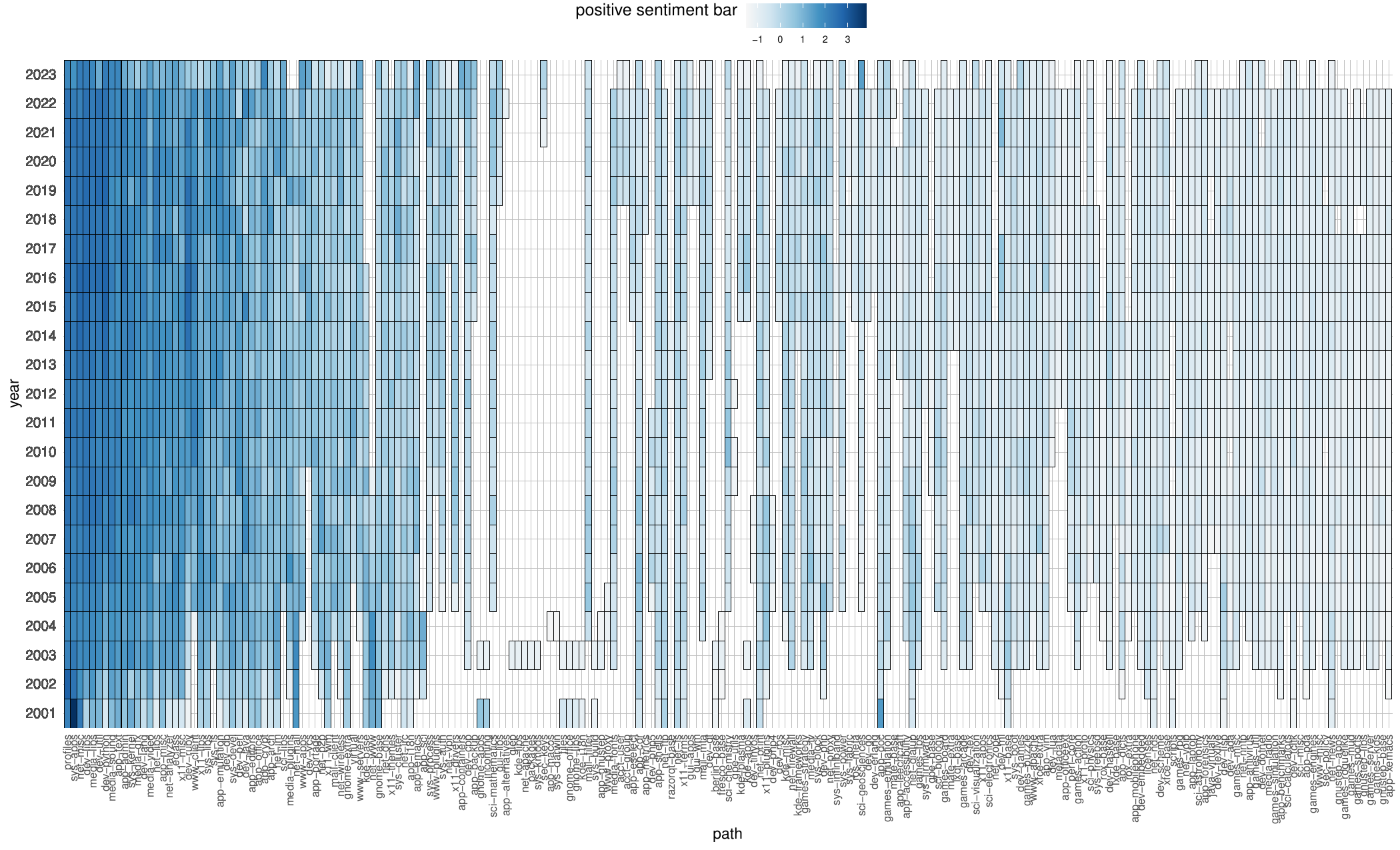}
    \caption{Heatmaps of number of positive  messages containing positive sentences by paths with standard normalisation applied yearly} 
\end{sidewaysfigure*}

\nocite{*}

\end{document}